\documentclass[useAMS,usenatbib]{mn2e}

\usepackage{times}
\usepackage{graphicx}

\def\pcm3{{\rm\thinspace cm^{-3}}}

\def\contcaption{\@conttrue\SFB@caption\@captype}

\def\n_h{{\rm n_{H}}}

\def\NH1{{$N_{\rm HI}~$}}

\def\ga{{\rm\thinspace gauss}}
\def\K{{\rm\thinspace K}}

\def\approxlt{\mathrel{\hbox{\rlap{\lower .5ex \hbox {$\sim$}}
        \raise .15 ex \hbox{$<$}}}}
\def\approxgt{\mathrel{\hbox{\rlap{\lower .5ex \hbox {$\sim$}}
        \raise .15 ex \hbox{$>$}}}}

\def\la{\mathrel{\hbox{\rlap{\hbox{\lower4pt\hbox{$\sim$}}}\hbox{$<$}}}}
\def\ga{\mathrel{\hbox{\rlap{\hbox{\lower4pt\hbox{$\sim$}}}\hbox{$>$}}}}
\newbox\grsign \setbox\grsign=\hbox{$>$} \newdimen\grdimen
\grdimen=\ht\grsign
\newbox\simlessbox \newbox\simgreatbox \newbox\simpropbox
\setbox\simgreatbox=\hbox{\raise.5ex\hbox{$>$}\llap
     {\lower.5ex\hbox{$\sim$}}}\ht1=\grdimen\dp1=0pt
\setbox\simlessbox=\hbox{\raise.5ex\hbox{$<$}\llap
     {\lower.5ex\hbox{$\sim$}}}\ht2=\grdimen\dp2=0pt
\setbox\simpropbox=\hbox{\raise.5ex\hbox{$\propto$}\llap
     {\lower.5ex\hbox{$\sim$}}}\ht2=\grdimen\dp2=0pt
\def\simgreat{\mathrel{\copy\simgreatbox}}
\def\simless{\mathrel{\copy\simlessbox}}

\title[SMC red giant stars - metallicities]{Red Giants in the Small Magellanic Cloud. II. Metallicity Gradient and Age-Metallicity Relation.}

\author[Dobbie et al.]{P.~D.~Dobbie$^{1}$\thanks{E-mail: paul.dobbie@utas.edu.au}, A.~A.~Cole$^{1}$, A.~Subramaniam$^{2}$, S. Keller$^{3}$\newauthor  \\
$^{1}$School of Physical Sciences, University of Tasmania, Hobart, TAS, 7001, Australia\\
$^{2}$Indian Institute of Astrophysics, Bengaluru 560034, India \\
$^{3}$Research School of Astronomy and Astrophysics, Australian National University, Canberra, Australia \\
}

\begin{document}

\date{Accepted . Received ; in original form }

\pagerange{\pageref{firstpage}--\pageref{lastpage}} \pubyear{2009}

\maketitle

\label{firstpage}

\begin{abstract}
We present results from the largest CaII triplet line metallicity study of Small Magellanic Cloud (SMC) field red giant stars 
to date, involving 3037 objects spread across approximately 37.5 deg$^{2}$, centred on this galaxy. We find a median metallicity 
of [Fe/H]=-0.99$\pm$0.01, with clear evidence for an abundance gradient of -0.075$\pm$0.011 dex deg$^{-1}$ over the inner 5$^{\circ}$. 
We interpret the abundance gradient to be the result of an increasing fraction of young stars with decreasing galactocentric radius,
coupled with a uniform global age-metallicity relation. We also demonstrate that the age-metallicity relation for an intermediate 
age population located 10kpc in front of the NE
of the Cloud is indistinguishable from that of the main body of the galaxy, supporting a prior conjecture that this is a stellar 
analogue of the Magellanic Bridge. The metal poor and metal rich quartiles of our RGB star sample (with complementary optical 
photometry from the Magellanic Clouds Photometric Survey) are predominantly older and younger than approximately 6Gyr, respectively.
Consequently, we draw a link between a kinematical signature, tentatively associated by us with a disk-like structure, and the 
upsurges in stellar genesis imprinted on the star formation history of the central regions of the SMC. We conclude that the increase 
in the star formation rate around 5-6Gyr ago was most likely triggered by an interaction between the SMC and LMC.

\end{abstract}

\begin{keywords}
galaxies: evolution - Local Group; galaxies: abundances; galaxies: individual: SMC; stars: abundances
\end{keywords}

\section{Introduction}

Low mass dwarfs are the most numerous galaxy type in the Universe and are found both in relative isolation and as satellites of larger assemblies. 
In the prevailing theoretical framework for galaxy formation, $\Lambda$-CDM \citep[e.g.][]{peebles03}, these small systems are the building blocks 
of the haloes of larger galaxies \cite{searle78}, a prediction that appears to be borne out by observations \citep[e.g.][]{frebel10}. Assemblies
comparable in mass to M31 and the Milky Way are anticipated to have cannibalised dozens of small systems over the last 10 Gyr \citep[][]{unavane96}, 
so it is important to develop a comprehensive picture of the chemical evolution of dwarf galaxies. This is initimately linked to the temporal and 
spatial progression of their star formation activity. In large assemblies like the Milky Way and M31, star formation appears to migrate outwards over 
time, in accord with $\Lambda$-CDM simulations which predict galaxy disks to continue growing as gas with progressively higher specific angular momentum 
accretes from within their dark matter haloes. In contrast, star formation in many dwarfs appears to have contracted inwards over time 
\citep[e.g.][]{wyder01,hidalgo09,zhang12} so these systems do not appear to behave merely as scaled down versions of large galaxies. 

Several physical processes have been identified which could result in the contraction of dwarf galaxies' star forming disks. These include hydro-dynamical
effects where the more rapid consumption of the gas by star formation in the partially pressure supported central regions of these galaxies leads to the 
inflow of material from larger radii to re-establish the balance, in turn depleting these parts of the fuel neccessary for further stellar genesis 
\citep[e.g.][]{stinson09, pilkington12, shen13}. The torques between stellar and gaseous structures that are induced in galaxy-galaxy mergers can remove 
angular momentum from the latter which in turn fall into the central regions of galaxies, resulting in increases in the star formation rates here 
\citep{mihos94,hopkins09}. Tidal interactions with neighbouring galaxies and ram-pressure stripping are efficient at removing gas, particularly from the 
peripheral regions of low mass galaxies \citep[e.g.][]{mayer06}. For example, in their recent detailed cosmological simulation, \cite{shen13} highlighted a
close encounter between two dwarfs which removed around 60\% of the gas from the lower mass system. Tidal forces can also promote the growth of bar 
instabilities \cite[][]{mayer01} which can lead to the radial inflow of gas and increases in the star formation rate in the inner regions of these galaxies. 

Tidal and ram-pressure stripping have also been invoked to explain, in part, the overprediction by $\Lambda$-CDM of the numbers of dwarf galaxies in the Local 
Volume, including the number that are satellites to the Milky Way \citep[][]{klypin99}. In the low luminosity regime the mass function of galaxies is expected 
to have a $M^{-1.9}$ form yet it is observed to be closer to $M^{-1}$ \citep{coles01}. Together with supernovae feedback, these mechanisms are anticipated to 
quench star formation by disrupting and expelling the gas from galaxies that they require to form stars, transforming the lowest mass dwarfs into very faint and 
dormant systems. Moreover, the cycle of stellar genesis and supernovae feedback is predicted to lead to distinctive bursts in the global star formation activity
of dwarf galaxies that can also help to resolve the disparity between the hypothetical centrally cusped dark matter profiles and the sigificantly flatter 
distributions inferred from empirical determinations of their rotation curves \citep[e.g.][]{governato10, deblok02}.

The two Magellanic Clouds are amongst the closest dwarf galaxies to the Sun and present an opportunity to study in some detail the role these various mechanisms play 
in regulating star formation and thus the chemical evolution of lower-luminosity systems. The Small Magellanic Cloud (SMC) is the smaller ($M$$\approx$2$\times$10$
^{9}$$M_{\odot}$), the more distant ($D$$\approx$60kpc) and the less studied of the pair. There is ample evidence that gas is being or has been relatively recently 
stripped from the SMC as a consequence of its interactions with the LMC and/or the Galaxy. For example, both Clouds are immersed within an extended body of 
diffuse HI gas that stretches out many tens of degrees across the sky, forming the Magellanic Stream and the Leading Arm \citep[e.g.][]{wannier72,mathewson74,putman03,nidever10}. \cite{fox13} have 
recently demonstrated this gas to have a composition consistent with that of the SMC 1.5-2.5Gyr ago, corresponding to the epoch during which this structure is believed 
to have formed. However, considering the proximity of the SMC and LMC to the Galaxy, they are relatively unusual in that they are gas rich whereas many dwarf galaxies 
within a few hundred kpc of the Milky Way and Messier 31 appear to be gas poor \citep[e.g.][]{grcevich09}. \cite{zaritsky00} have proposed that the SMC has recently 
accreted a gas cloud, in view of the highly fragmented distribution of its younger stellar populations and the stark difference between the spatial distributions of 
these and the older generations of stars. Different stellar populations in a number of other dwarf irregular galaxies, such as NGC\,6822 \citep[e.g.][]{letarte02} and
IC\,10 \citep{borissova00}, also appear to have distinctive distributions.

\begin{figure}
\includegraphics[angle=0, width=\linewidth]{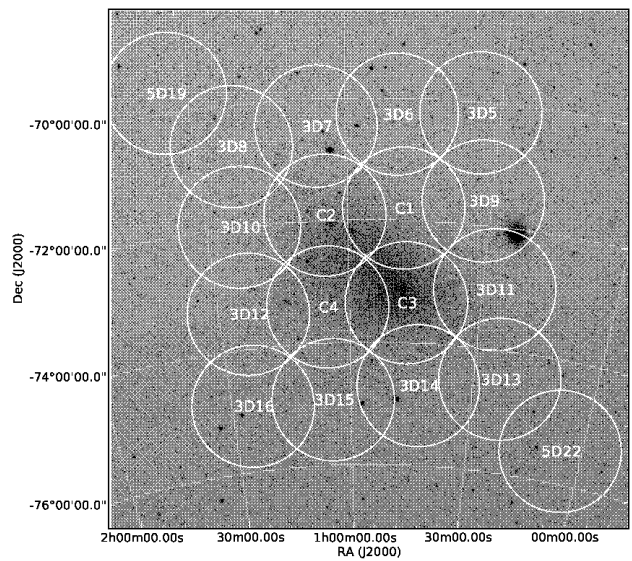}
\caption{A 9$^{\circ}$$\times$9$^{\circ}$ image of the sky centered on the SMC \citep[3.4$\mu$m WISE data, ][]{wright10}. The 
circles, which each corresponds to an AAT + 2dF/AAOmega pointing, highlight the areas included in our photometric and 
spectroscopic survey. Note that at least two distinct fibre configurations were observed for each of the four central fields.}
\label{smcmap}
\end{figure}

Within the last decade or so, there have been several detailed investigations of the star formation history of the SMC. For example, \cite{dolphin01} analysed 
a deep colour-magnitude diagram constructed from Hubble Space Telescope (HST) and Wide Field Planetary Camera (WFC) data for a field several degrees to the 
north-west of the Bar. They found a generally low level of star formation here, that peaked around 5-8Gyr ago and that has dropped off sharply in more recent 
times. Similarly, \cite{chiosi07} analysed HST imaging of three fields located at the southern end of the Bar. They unearthed a distinct peak in the star formation
rate around 3-6 Gyr ago and concluded that only very inefficient star formation proceeded in these three fields prior to this time. \cite{sabbi09}, \cite{cignoni12}
and \cite{cignoni13} have examined deep HST imaging of six further fields across the SMC and their results are accordant with those of the previous space-based 
studies. This prominent rise in the star formation rate around 5Gyr ago is also recovered in a ground-based imaging study of 12 fields widely distributed 
across the SMC \citep{noel09}. These investigators also find evidence of a global burst of star formation around 10 Gyr ago, which is not inferred from the HST
observations. \cite{harris04} similarly found there to have been a burst of star formation in the Cloud around this time based on their analysis of a much more 
spatially extensive but photometrically shallower dataset. However, it has been suggested that their conclusion regarding this early burst of star formation activity
is compromised by the lack of depth of their imaging data, which does not reach to below the oldest main sequence turn-off \citep[e.g.][]{subramanian12}. 

\begin{table*}
\begin{minipage}{160mm}
\begin{center}
\caption{Details of the 3037 SMC red-giant stars, including our CaII triplet equivalent width measurements and abundance estimates.}
\label{rgstars}

\begin{tabular}{|l|l|l|r|r|r|r|r|r|r|r|}
\hline
  \multicolumn{1}{|c|}{2MASS\,J} &
  \multicolumn{1}{c|}{RA} &
  \multicolumn{1}{c|}{Dec} &
  \multicolumn{1}{c|}{$J$} &
  \multicolumn{1}{c|}{$\delta$$J$} &
  \multicolumn{1}{c|}{$K_{s}$} &
  \multicolumn{1}{c|}{$\delta$$K_{s}$} &
  \multicolumn{1}{c|}{$\Sigma W$} &
  \multicolumn{1}{c|}{$\delta$$\Sigma W$} &
  \multicolumn{1}{c|}{[Fe/H]} &
  \multicolumn{1}{c|}{$\delta$[Fe/H]} \\

  \multicolumn{1}{|c|}{} &
  \multicolumn{1}{|c|}{hh:mm:ss.ss} &
  \multicolumn{1}{|c|}{ ${^\circ}$:$'$:$''$} &
  \multicolumn{1}{c|}{/mag.} &
  \multicolumn{1}{c|}{/mag.} &
  \multicolumn{1}{c|}{/mag.} &
  \multicolumn{1}{c|}{/mag.} &
  \multicolumn{1}{c|}{/\AA} &
  \multicolumn{1}{c|}{/\AA} &
  \multicolumn{1}{c|}{} &
  \multicolumn{1}{c|}{} \\

\hline

  00:00:28.28 & -75:33:04.4 & 00002828-7533044 &  14.68 &   0.04 &  13.72 &   0.05 &   7.11 &   0.30 &  -0.97 &   0.12\\
  00:02:31.86 & -75:12:37.1 & 00023186-7512371 &  14.26 &   0.03 &  13.43 &   0.03 &   6.48 &   0.30 &  -1.22 &   0.11\\
  00:04:35.32 & -75:48:42.1 & 00043532-7548421 &  14.10 &   0.03 &  13.15 &   0.03 &   8.04 &   0.34 &  -0.75 &   0.13\\
  00:05:35.96 & -75:27:07.7 & 00053596-7527077 &  14.41 &   0.03 &  13.53 &   0.05 &   7.25 &   0.24 &  -0.95 &   0.10\\
  00:06:26.25 & -75:23:45.0 & 00062625-7523450 &  14.67 &   0.04 &  13.79 &   0.04 &   6.16 &   0.36 &  -1.27 &   0.13\\

\hline
\end{tabular}
\end{center}
\end{minipage}
\end{table*}

There have also been a number of spectroscopic studies of the intermediate age field star population of the SMC aimed at probing its star formation history. 
For example, \cite{carrera08} observed 350 RGB stars in 13 pointings scattered within 4-5$^{\circ}$ of the center of the Cloud and detected a radial metallicity 
gradient in this population. Drawing from their analysis of the form of the age-metallicity relation, they attributed this to the younger, more metal rich objects
being the most centrally concentrated component of the RGB star population. The results from the deeper imaging studies and the \citeauthor{carrera08} work are 
indicative of an outside to in migration of star formation activity in the SMC. Intriguingly though, in another, more recent, spectroscopic study of 360 field red 
giant stars drawn from 15 distinct pointings extending out to 7-8$^{\circ}$, \cite{parisi10} found no compelling evidence of a metallicity gradient in the 
intermediate age stellar population.
The most comprehensive study to date of the metallicities and the ages of the SMC open clusters reached a similar conclusion \citep{piatti12}. Clearly, to build 
a better understanding of the progression of star formation activity throughout the SMC, it is first necessary to confirm or otherwise the existence of a radial 
variation in 
the metallicity of the intermediate age population. Additionally, to gain greater insight into the drivers of star formation activity in the SMC, and dwarf galaxies 
in general, it is informative to search for and identify any associations between the ages and kinematics of the RGB population and variations in the Cloud's historical
 star 
formation rate.

In the present work we describe the analysis and the interpretation of stellar metallicity measurements for 3037 red-giant stars located across 37.5 deg$^{2}$ 
of the SMC. Our study of the kinematics of these stars is presented elsewhere (Dobbie et al. 2014). In the following sections we discuss our measurements of the stellar 
abundances and our analysis of these values. After cross-matching our sources with the Magellanic Clouds Photometric Survey \citep[MCPS, ][]{zaritsky02} to obtain 
optical 
photometry, we use a published relation between [Fe/H], $V$ and $V$-$I$ to derive their ages. We compare our age-metallicity relations for the main body of the SMC
and a population of red-giants 10 kpc in front of the NE of the Cloud recently proposed to be the stellar analogue of the Magellanic Bridge. Next we examine the 
relative age distributions of the most metal rich and metal poor quartiles of the RGB population and examine their kinematics and relative ages in the context of the 
historical star formation rates of the SMC. We conclude the paper with a summary of our main findings.

\section[]{SMC RGB stars}

\subsection{Candidate selection and spectroscopic follow-up}

\label{survey}

Although the details of our selection and optical spectroscopic follow-up of SMC red giant stars are described in Dobbie et al. (2014), for 
completeness we provide a recap of the main points of this aspect of our work here. Candidate SMC red-giants were selected from their location
in a $J$,$J$--$K_{S}$ colour-magnitude diagram, constructed for sources with photometric uncertainties of less than 0.5 mag. in both bands, using 
the near-IR photometry of the 2 micron All-Sky Survey \cite[2MASS; ][]{skrutskie06} point source catalogue (PSC). Our survey covered an area of 
approximately 37.5 deg$^{2}$, centred on the Cloud (figure~\ref{smcmap}). We selected sources to the red of a line defined by $J$ = 26.5 - 20
$\times$ ($J$--$K_{S}$) and blueward of $J$--$K_{S}$=2.0 or $J$--$K_{S}$=1.25 for 12.0$\le$$J$$ <$13.9 and 13.9$\le$$J$$\le$15.2, respectively. We 
eliminated from this initial list, sources that were flagged as possible blends, as having photometry contaminated by image artifacts or nearby 
bright objects and/or as lying within the boundaries of catalogued extended objects. The resulting preliminary catalogue of 92\,893 objects includes
both SMC red-giant branch (RGB) and asymptotic-giant branch (AGB) stars.

Spectroscopic follow-up data for approximately 7000 of these targets were acquired during the period 18-21 October 2011, with the 2dF/AAOmega fibre-fed 
multi-object optical-spectrograph \citep{saunders04,sharp06} and the 4.2m Anglo-Australian Telescope (AAT). The blue and red arms of AAOmega were 
configured with the 1500V (R$\approx$4000) and 1700D (R$\approx$10000) gratings and tuned to central wavelengths of 5350\AA\ and 8670\AA, respectively. We 
utilised 23 different field configurations, the centers and the exposure times for which are reported in Table~1 of Dobbie et al. (2014). The data were 
reduced using the Australian Astronomical Observatory's {\tt 2dFDR} pipeline \citep{bailey98,sharp10}. The final spectrum for each source was obtained by 
combining the data from the two to four individual exposures obtained per plate configuration. 

The red-arm spectra were normalised with low-order polynomials the forms of which were determined by matching the data to a multiplicative combination of 
these and normalised synthetic spectrum drawn from the library of \cite{kirby11}. Radial velocities for the sources, in most cases accurate to better than 
5 kms$^{-1}$, were obtained by cross-correlating the normalised spectra with AAOmega datasets for several RGB stars drawn from the clusters NGC\,362, Melotte\,66
and NGC\,288, for which reliable radial velocities were available in the literature. This process was also undertaken for the C-stars in the sample, using AAOmega 
observations of six C-rich giants taken from the study of \cite{kunkel97} as the radial velocity templates.
Subsequently, we utilised the blue-arm spectra to eliminate contaminating field dwarfs from the sample and to resolve the K-type from the C-rich giants. We 
constructed orthogonal basis vectors to describe a 50\AA\ wide section of these data centered on $\lambda$$\approx$5170\AA\ which covered the $\lambda$5167, 5172 and 
5183\AA\ Mg b lines and the distinctive C Swan band feature at 5165\AA. After first removing 449 C-rich stars from the spectroscopic sample, we selected 4\,172 objects, 
4\,077 of which are unique, with radial velocities in the range 50$\le$v$_{r}$$\le$250kms$^{-1}$ (and lying to the right of the line defined by PC$_{1}$= -0.013 v$_{r}$ 
+ 1.0, see Dobbie et al. 2014 for details) as probable K-type red giant members of the SMC. Finally, as the RGB tip in the SMC is found to lie at $J$$\approx$13.7 
\citep{cioni00}, we conservatively cut at $J$$\ge$14 to define a sample of 3\,037 sources we anticipate to be dominanted by SMC RGB stars. Details of these
objects, including metallicity estimates, are listed in Table~\ref{rgstars}.

\subsection{RGB star metallicity measurements}

\begin{figure}
\label{fehhist}
\includegraphics[angle=0, width=\linewidth]{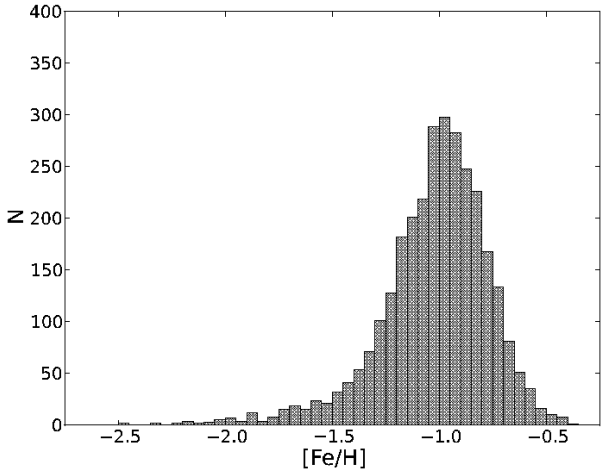}
\caption{The histogram of our metallicity ([Fe/H]) estimates for the entire spectroscopically observed sample of SMC RGB stars.}
\label{fehhist}
\end{figure}

%{\it Expand details of how equivalent widths were measured here, how uncertainties were derived and checks on repeatability.}
To gauge the metallicities of the SMC RGB stars we have applied the widely documented technique involving the summation ($\Sigma W$) of the 
equivalent widths of the $\lambda$8498, 8542 and 8662\AA\ CaII lines \citep{olszewski91,armandroff91,cole00}. As the 2MASS survey \citep{cutri03} 
currently represents the most uniform and reliable photometric database for our spatially extensive sample, to obtain estimates on the scale of \cite{carretta97},
we have followed the approach of \cite{warren09}. We adopted as proxies for the luminosities of the stars, the differences between their 2MASS $K_{s}$ 
magnitudes and that of their host population's red clump. Importantly, \cite{pietrzynski03} find that in the $K_{s}$ band, the absolute magnitude of the
red clump displays little dependence on a population's age and metallicity within relatively broad ranges, $2\simless \tau$ (Gyr) $\simless9$ and $-1.8<$[Fe/H]$<-0.5$, 
respectively. Variations in the SMC population's age and metallicity slightly outwith these ranges (e.g. $\tau\approx1--2$ (Gyr)) are also expected to have only
a minor impact on this work, $\Delta$[Fe/H]$\simless$0.03 \cite[e.g.][]{grocholski02, warren09}.
The characteristics of the CaII lines 
have been determined by matching them to Voigt profiles at the stellar velocities as measured by the cross-correlation procedure. To assess the
uncertainties in the line equivalent width measurements, we utilised a bootstrap with replacement approach, whereby for each spectrum, we generated several hundred resampled versions and refitted the line profiles in every one.
The resulting uncertainty estimates have been examined with repeat observations of approximately 100 SMC stars and have been found to be approximately 
30-40\% smaller than the scatter between the independent sets of measurements. There was no apparent correlation between the size of this disparity and 
object magnitude so the equivalent width uncertainties were simply scaled-up accordingly and adopted for the remainder of this work. 

\begin{figure}
\label{fehspat}
\includegraphics[angle=0, width=\linewidth]{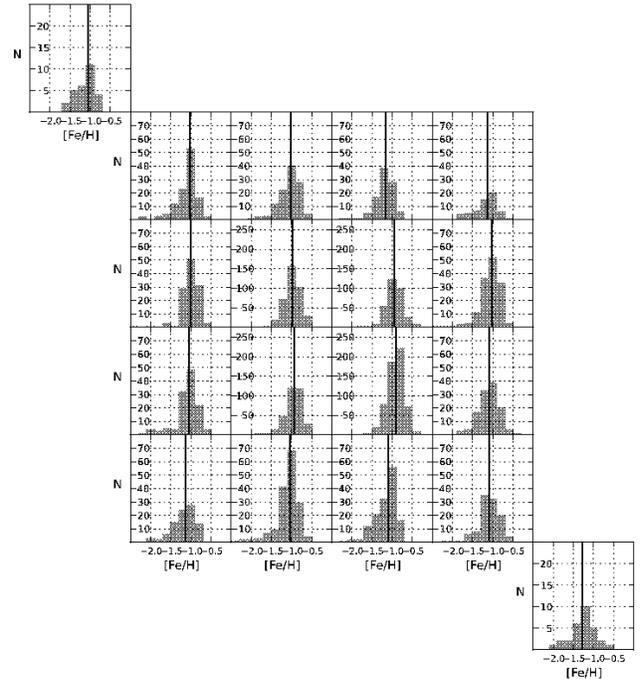}
\caption{Histograms for the metallicities, [Fe/H], of the spatial sub-samples of SMC RGB 
stars. The median metallicity of each sub-sample is highlighted (solid black line).}
\label{fehspat}
\end{figure}

To appraise our techniques, the RGB members of four star clusters of known metallicity were also observed during our four night campaign 
at the AAT. These were drawn from the globular clusters M\,30, NGC\,362 and NGC\,104 and the open cluster Melotte 66 and collectively span
a relatively broad range of metallicity, -2.0 $\simless$ [Fe/H] $\simless$ -0.5. The red-clump magnitudes ($K_{S_{RC}}$) adopted for these populations 
were taken from the literature \citep[e.g.][]{warren09} or from the median $K_{S}$ magnitude of the fundamental-mode RR Lyrae members of their 
horizontal branches for which 2MASS measurements are available (Table~\ref{RCM}). Subsequently, for each of these four populations a second order 
polynomial has been fitted to the $\Sigma W$ determinations as a function of luminosity. We have examined the behaviour of this relation and found 
no compelling evidence in any of these four datasets for significant departures from a simple linear trend. The weighted mean of the four function 
gradients, $\beta_{\K_{\rm S}}$=0.47$\pm$0.05, appears to be in excellent agreement with the result from our prior study of 17 cluster populations 
\citep[$\beta_{\K_{\rm S}}$=0.48$\pm$0.01; ][]{warren09}. Indeed, adopting the gradient ($\beta_{\K_{\rm S}}$) and the relation between $\Sigma W^{\prime}$ 
and [Fe/H], [Fe/H]=(-2.738$\pm$0.063)+(0.330$\pm$0.009)\,$\Sigma W^{\prime}$, from this earlier work, we determine metallicities for  M\,30, 
NGC\,362 and NGC\,104 and the open cluster Melotte 66 of [Fe/H] = -1.91$\pm$0.07, -1.13$\pm$0.08, -0.75$\pm$0.07 and -0.47$\pm$0.09, respectively.
These are in accord with estimates in the literature \citep[e.g. \protect{[Fe/H]} = -1.91$\pm$0.00, -1.15, -0.70$\pm$0.03 and 
-0.48$\pm$0.06; ][]{carretta97,cole04} and suggest that the data reduction and analysis processes applied here do not introduce substantial systematic
error in our abundances, at least with respect to our prior work.

\begin{table}
\begin{minipage}{53mm}
\begin{center}
\caption{Adopted $K_{S}$ magnitudes for the red clumps in the four fiducial clusters.}
\label{RCM}
\begin{tabular}{lcc}
\hline
Cluster & $K_{S_{RC}}$ & Reference \\ 
      & (/mag.) &  \\
\hline

M\,30 & 13.83$\pm$0.03 &  1 \\
NGC\,362 & 14.28$\pm$0.16 & 1  \\
NGC\,104 & 11.94$\pm$0.12 & 2  \\
Melotte 66 & 11.74$\pm$0.13 & 2\\

\hline
\end{tabular}
\end{center}
1. This work
2. \cite{warren09}
\end{minipage}
\end{table}

\begin{figure}
\label{fehlum}
\includegraphics[angle=0, width=\linewidth]{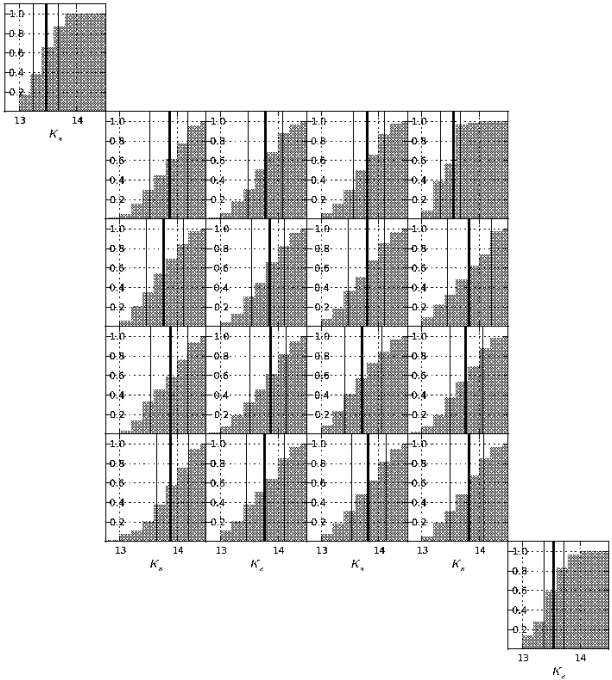}
\caption{Cumulative $K_{S}$ magnitude distributions for our spatial sub-samples of SMC RGB stars. The median (thick black line) and lower and 
upper quartiles (black lines) of the distributions are overplotted.}
\label{magspat}
\end{figure}

\begin{table*}
\begin{minipage}{160mm}
\begin{center}
\caption{A table summarising the results from examining the metallicity histograms for our spatial 
sub-samples of RGB stars. The median, the 25th and the 75th percentiles of the $K_{s}$ magnitude distribution of each sub-sample 
are also listed.}
\label{Metals}
\begin{tabular}{lrccc}
\hline
Field & N$_{\rm tot}$ & $r$ & $\tilde{\rm [Fe/H]}$(Q$_{1}$,Q$_{3}$) & $\tilde{K_{s}}$ (Q$_{1}$,Q$_{3}$) \\ 
    & &  /$^{\circ}$ &    & /mag. \\ 
\hline

C1     & 321 & 1.28 &-0.93$\pm$0.02 (-0.81,-1.07)  & 13.80 (13.48, 14.10)  \\
C2     & 390 & 1.34 &-0.97$\pm$0.02 (-0.86,-1.10)  & 13.85 (13.53, 14.12) \\
C3     & 586 & 0.86 &-0.89$\pm$0.01 (-0.77,-1.03)  & 13.71 (13.41, 14.06)  \\
C4     & 333 & 1.26 &-0.93$\pm$0.02 (-0.82,-1.07)  & 13.87 (13.51, 14.15) \\
3D05   & 57 &  4.29 &-1.12$\pm$0.05 (-0.96,-1.36)  & 13.56 (13.32, 13.67)  \\
3D06   & 99 &  3.48 &-1.17$\pm$0.03 (-1.06,-1.32)  & 13.81 (13.57, 14.10)  \\
3D07   & 112 & 2.89 &-1.01$\pm$0.03 (-0.87,-1.20)  & 13.78 (13.48, 14.07) \\
3D08   & 115 & 3.21 &-1.03$\pm$0.03 (-0.94,-1.21)  & 13.86 (13.53, 14.19)  \\
3D09   & 145 & 2.59 &-1.02$\pm$0.03 (-0.90,-1.21)  & 13.82 (13.44, 14.21)  \\
3D10   & 120 & 2.56 &-1.00$\pm$0.03 (-0.88,-1.14)  & 13.77 (13.45, 14.11)  \\
3D11   & 121 & 1.79 &-1.09$\pm$0.03 (-0.93,-1.24)  & 13.76 (13.50, 14.06)  \\
3D12   & 124 & 3.01 &-1.05$\pm$0.04 (-0.92,-1.16)  & 13.88 (13.53, 14.18)  \\
3D13   & 108 & 2.46 &-1.08$\pm$0.03 (-0.93,-1.23)  & 13.82 (13.49, 14.08)  \\
3D14   & 142 & 2.31 &-1.10$\pm$0.03 (-0.97,-1.31)  & 13.81 (13.53, 14.14)  \\
3D15   & 170 & 3.03 &-1.04$\pm$0.04 (-0.93,-1.21)  & 13.77 (13.45, 14.12)  \\
3D16   & 92 &  4.39 &-1.13$\pm$0.04 (-0.95,-1.31)  & 13.88 (13.65, 14.20)  \\
5D19   & 29 &  4.43 &-1.06$\pm$0.08 (-0.95,-1.31)  & 13.47 (13.25, 13.68)  \\
5D22   & 29 &  4.07 &-1.27$\pm$0.07 (-1.10,-1.44)  & 13.55 (13.37, 13.73)  \\

\hline
\label{Metals}
\end{tabular}
\end{center}
\end{minipage}
\end{table*}

\section{Metallicities}

\subsection{The overall distribution of [Fe/H]}

We have adopted a singular value of $K_{S_{RC}}$=17.35$\pm$0.02 for the magnitude of the SMC red clump, since deep and uniform 
photometry of our extensive survey area, suitable for assessing its dependence on RA and declination, is not yet available.  
This value is based on deep $J$ and $K$ imaging for 10 pointings (totalling an area of approximately 0.017 deg$^{2}$) towards
the Cloud with the ESO/NTT telescope and the SOFI instrument, that was used to isolate the giant branch and locate the peak in the $K$ 
band luminosity function of this population \citep{pietrzynski03}. It should represent a fair estimate of the magnitude zeropoint 
for our abundance analysis since the $U$, $B$, $V$ and $I$ band MCPS \citep[][]{zaritsky02} investigation of extinction in the direction
of the central 4.5$^{\circ}$$\times$4$^{\circ}$ degrees of the SMC, has found no evidence of strong differential reddening towards the 
intermediate and older age populations. This conclusion has been corroborated by more recent studies of these stellar generations of the
SMC, although three small regions of somewhat higher extinction have been identified within the Bar and
the Wing \citep[e.g. $E(V-I)=0.16$, ][]{haschke11,subramanian12}. As A$_{K}$/E($B$-$V$)=0.36 \citep[for R=3.1, ][]{fitzpatrick99}, these
comparably subtle variations should have a negligible impact on our metallicity estimates. However, we take them into account in 
estimating the ages of our RGB stars in Section~\ref{s6} since this step is reliant on $V$ and $I$ band photometry.

\cite{cioni00} determined the tip of the RGB in the SMC to lie at $J$$\approx$13.7 so, conservatively, the 3\,037 stars with $J$$\ge$14, 
which weren't flagged as C-rich in Section~\ref{survey} have been selected for further scrutiny. For all these objects, we have calculated 
$K_{S}$-$K_{S_{RC}}$ and their reduced equivalent widths, $\Sigma$ W$^{\prime}$, assuming $\beta_{\K_{\rm S}}$=0.48$\pm$0.01 \citep{warren09}.  
Subsequently, relying again on results from our earlier work, we have calculated their metallicities, [Fe/H]. A histogram of the results 
of this process is shown in figure~\ref{fehhist}. The overall shape of this distribution is comparable to that seen in previous metallicity 
studies of SMC RGB field stars \citep[e.g.][]{parisi10}. There is a relatively rapid decline in the number of objects towards higher metallicities,
with a negligible number found to have [Fe/H]$\simgreat$-0.4. However, there is a more gradual tailing off in the number of stars towards lower 
metallicities, -2.5 $<$ [Fe/H] $<$ -2.0. This is reflected by the negative skewness calculated for the observed distribution ($\gamma_{1}$=-1.819$
\pm$0.044). An estimate of this sample's kurtosis ($\gamma_{2}$=7.900$\pm$0.089) and the results of a Kolmogorov-Smirnov test for normality 
(P=3e-19) also indicate that a simple Gaussian is not a particularly good representation of the underlying metallicity distribution. The
histogram resembles that derived for the RGB stars in the bar of the Large Magellanic Cloud \citep[LMC, ][]{cole04}, albeit shifted to lower
metallicities. Recognising the departure of this distribution from normality, rather than determining a mean metallicity, we have calculated 
the median value for these stars, [Fe/H]$\approx$-0.993$\pm$0.006. This is comparable to the median metallicity of the RGB stars investigated 
by \cite{parisi10} (Fe/H$\approx$-0.985$\pm$0.020).

\subsection{Radial variation in [Fe/H]}
\label{spatial}

To investigate how the metallicities of our stars depend \citep[e.g.][]{carrera08}, or otherwise \citep[e.g.][]{parisi10}, on their projected 
location in the SMC, first we have split the sample of 3037 objects approximately in half at a galacto-centric distance of around 1.7$^{\circ}$. 
With both sub-samples we have examined metallicity as a function of position-angle and in each case found no evidence of an angular dependence. 
This is accordant with prior results on the low inclined orientation of the intermediate age population in the SMC \citep[e.g.][]{subramanian12}.
Next, the 3\,093 RGB star metallicity measurements (ie. including objects observed twice with different field configurations) have been split 
into 18 spatial sub-samples, which correspond to our 2dF field pointings. The largest and the smallest of these contains 589 
and 29 RGB stars, respectively. Basic parameters for the metallicity distribution of each sub-sample have been obtained by constructing a [Fe/H]
histogram and calculating the median metallicity (figure~\ref{fehspat}). The output from this process is listed in Table~\ref{Metals}. An examination 
of figure~\ref{fehspat}, reveals that, ostensibly, the spectroscopically observed stars located in the inner regions of our survey area tend 
to be more metal rich than those at larger radial distances (r$>$2$^{\circ}$) from the optical center of the Cloud.

\begin{figure}
\label{fehlum}
\includegraphics[angle=0, width=\linewidth]{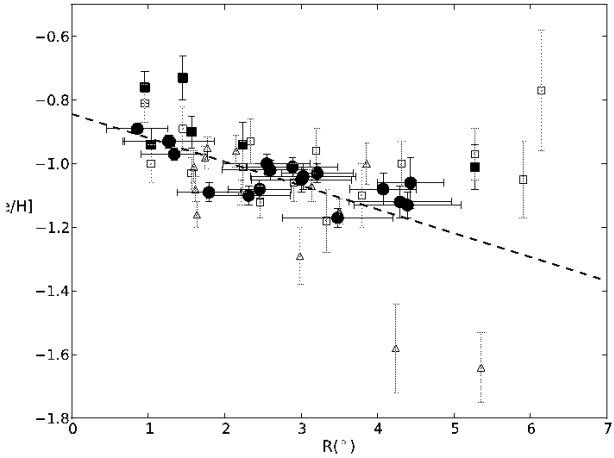}
\caption{Our determinations of the median metallicity of RGB stars as a function of projected radius from the optical center of the SMC (filled black circles). The results from two previous studies of smaller samples of RGB stars are overplotted, \citet[][open triangles]{carrera08} and \citet[][open squares]{parisi10}. Our improved measurements indicate that the metallicity decreases with increasing distance from the Cloud center, at least out to R$\approx$5$^{\circ}$, but not as steeply as found by \citet[][]{carrera08}. Revised metallicity estimates for a number of fields from \citet[][]{parisi10} obtained using data from the IRSF Magellanic Clouds Point Source Catalogue \citep[filled black squares, ][]{kato07} are overplotted and are largely consistent with the trend de-lineated by our new measurements.} 
\label{metrad}
\end{figure}

\begin{figure}
\label{metmap}
\includegraphics[angle=0, width=\linewidth]{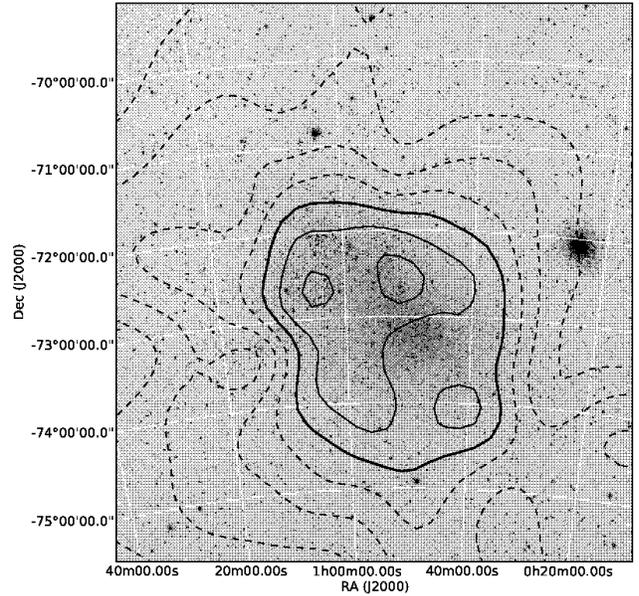}
\caption{A 6.0$^{\circ}$$\times$6.5$^{\circ}$ contour map of the RGB star metallicities. The data around each grid point were smoothed using an adaptive gaussian kernel 
with a width corresponding to one third of the distance to the 200th closest star. The contours correspond to steps of 0.05 dex in [Fe/H] (heavy line [Fe/H]=-1.0 and 
the dashed lines have lower values).}
\label{metmap}
\end{figure}

\begin{figure}
\label{metmapkern}
\includegraphics[angle=0, width=\linewidth]{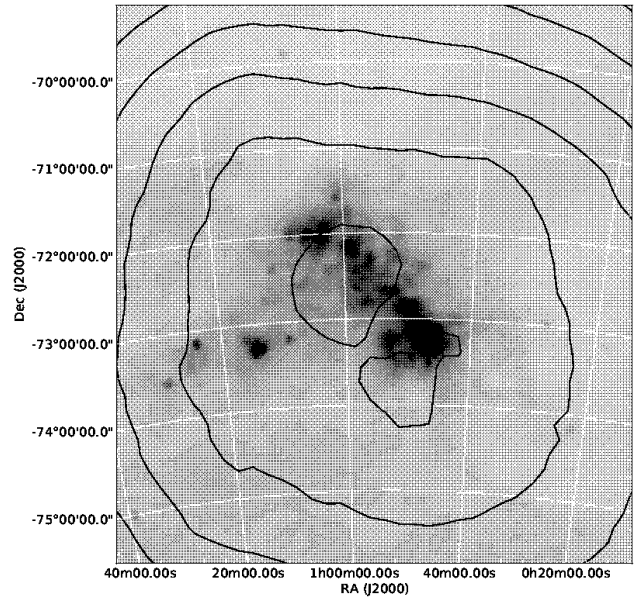}
\caption{A 6.0$^{\circ}$$\times$6.5$^{\circ}$  map of smoothing kernel width used in producing figure~\ref{metmap}. The contours correspond to steps of 10 arcmin with the innermost contour corresponding to 10 arcmin.}
\label{metmapkern}
\end{figure}

To assess the reality of the apparent radial variations in our measurements we have examined a number of potential sources of systematic error in these 
results. Firstly, following the argument made above, it is improbable that differences of this magnitude can be attributed to differential
reddening. Indeed, even the total neglect of much stronger, patchier extinction, as is associated with the early-type stellar population of the SMC 
(A$_{V}$$\simless$2), would for R$\approx$3.1, only account for $\delta$[Fe/H]$\approx$0.05 \citep{zaritsky02}. Secondly, we have explored the 
magnitude distributions of the stars in each of our sub-regions in case our spectroscopic follow-up was biased towards proportionally more bright 
or more faint stars depending on location in the Cloud. It is conceivable that, coupled to a subtle systematic uncertainty in the $\Sigma$ W, $K_{S}$-$
K_{S_{RC}}$ relation which was applied to derive the reduced equivalent widths, this resulted in the observed field-to-field variations. However, the 
cumulative distributions of the $K_{s}$ magnitudes of the targets in the sub-regions do not support this notion (figure~\ref{magspat}). There are 
only relatively small differences in the median magnitude from field-to-field and the weighted Pearson product-moment correlation coefficient for
the 18 measurement pairings of mean [Fe/H] and median $K_{S}$ is only r=0.35, consistent with the lack of a significant trend between these two 
variables. Thirdly, we have considered the impact of variations in the mean distance of the intermediate-age stellar population 
across the face of the SMC. A recent photometric analysis of the distance distribution of red clump stars $4^{\circ}$ from the optical center reveals a 
substantial population located about 10kpc in front of the main body of the Cloud, to the north and east. Other investigations of red clump and RR Lyrae 
stars have shown that in the central regions of our survey area, the SMC extends along a path length of up to 14kpc \citep{subramanian12}. Potentially,
 we might have overestimated the luminosities of individual stars, at least around the northern and the eastern limits of our survey region, by 0.3-0.4 mag.
 However, this would lead to systematic errors in their metallicity determinations of only $\delta$[Fe/H]$\approx$0.05. Moreover, as the lower RGB is more 
heavily populated than the upper RGB due to the different evolutionary timescales, our initial photometric selection criteria in conjunction with the large
 line of sight depth should have resulted in the loss from our central survey region of more stars where the metallicity is overestimated as opposed to 
underestimated. Thus it is unlikely that varying distances or depths of the sub-samples across the Cloud are responsible for the observed variations in
 metallicity.

\begin{table*}
\begin{minipage}{160mm}
\begin{center}
\caption{Details derived from the 1865 RGB star sources (1835 of which are unique) for which both $V$ and $I$ band photometry is available in the MCPS 
and extinction levels can be estimated from the work of {\protect \cite{haschke11}}.}
\label{ages}
\begin{tabular}{|l|l|l|r|r|r|r|r|r|r|r|}
\hline
  
  \multicolumn{1}{c|}{RA} &
  \multicolumn{1}{c|}{DEC} &
  \multicolumn{1}{|c|}{2MASS\,J} &
  \multicolumn{1}{c|}{$V$} &
  \multicolumn{1}{c|}{$\delta$ $V$} &
  \multicolumn{1}{c|}{$A_{V}$} &
  \multicolumn{1}{c|}{$I$} &
  \multicolumn{1}{c|}{$\delta$ $I$} &
  \multicolumn{1}{c|}{$A_{I}$} &
  \multicolumn{1}{c|}{$\log{Age}$} &
  \multicolumn{1}{c|}{$\delta\log{Age}$} \\

  \multicolumn{1}{|c|}{hh:mm:ss.ss} &
  \multicolumn{1}{|c|}{ ${^\circ}$:$'$:$''$} &
  \multicolumn{1}{|c|}{} &
  \multicolumn{1}{c|}{/mag.} &
  \multicolumn{1}{c|}{/mag.} &
  \multicolumn{1}{c|}{/mag.} &
  \multicolumn{1}{c|}{/mag.} &
  \multicolumn{1}{c|}{/mag.} &
  \multicolumn{1}{c|}{/mag.} &
  \multicolumn{1}{c|}{/$\log(Gyr)$} &
  \multicolumn{1}{c|}{/$\log(Gyr)$} \\

\hline

  00:23:56.14 & -73:59:26.3 & 00235613-7359263 &  16.51 &   0.03 &   0.07 &  15.03 &   0.03 &   0.04 &  10.13 &   0.00\\
  00:24:11.97 & -74:54:29.9 & 00241196-7454299 &  16.94 &   0.23 &   0.07 &  15.46 &   0.07 &   0.04 &   9.81 &   0.68\\
  00:24:13.44 & -74:30:06.3 & 00241344-7430063 &  17.24 &   0.04 &   0.07 &  15.84 &   0.04 &   0.04 &   9.93 &   0.32\\
  00:24:40.09 & -73:43:46.0 & 00244009-7343460 &  17.10 &   0.03 &   0.05 &  15.74 &   0.02 &   0.03 &  10.13 &   0.00\\
  00:24:42.47 & -73:45:31.2 & 00244247-7345312 &  16.77 &   0.22 &   0.05 &  15.53 &   0.02 &   0.03 &   9.53 &   0.80\\

\hline\end{tabular}
\end{center}
\end{minipage}
\end{table*}

Accepting that the field-to-field differences seen in the median metallicities of the SMC RGB stars are real, we have examined in more detail 
how [Fe/H] changes with galacto-centric distance. For consistency with prior metallicity studies \citep[e.g.][]{piatti07, parisi10}, the elongation of 
the structure of the Cloud has been accounted for by working within an elliptical co-ordinate system. This has its origin at the optical 
center of the SMC and has its major-axis closely aligned with the SMC bar. We have accepted the semi-major axis of an ellipse on which a star 
lies as a proxy for its galacto-centric distance. In lieu of the results of a recent detailed study of the three-dimensional distribution of 
the old and intermediate-age stellar populations of the SMC \citep{subramanian12}, we have adopted a major-axis position angle of 55.3$^{\circ}$,
east of north, and have assumed the ratio of the semi-major to semi-minor ($b$) axes to be $a/b=1.5$. The radial distances to our sub-samples were 
taken to be the means of those of all RGB stars within the corresponding 2dF field pointings and are shown in Table~\ref{Metals}. It can
be seen that our measurements reach out to 4--5$^{\circ}$, comparable to the radial extent of the \cite{carrera08} study.

\begin{figure}
\label{agemetpt}
\includegraphics[angle=0, width=\linewidth]{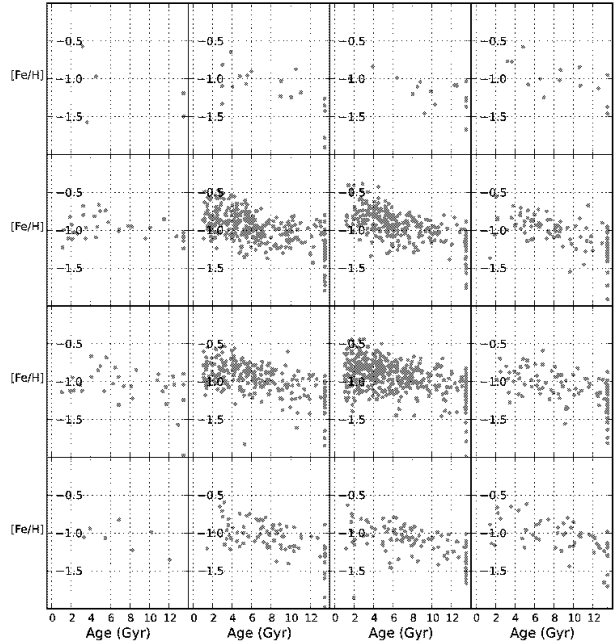}
\caption{A plot of [Fe/H] as a function of age for the 1865 RGB stars with $V$ and $I$ photometry as distributed across each of our 
inner 16 2dF field pointings.}
\label{agemetpt}
\end{figure}

Our new estimates of [Fe/H] for the SMC field RGB population convey a negative correlation between metallicity and distance from the optical center
of the Cloud (figure~\ref{metrad}, filled black circles). While there is greater scatter in the measurements at larger distances (this is likely at 
least partly due to the smaller numbers of stars in the outermost sub-samples), a $\chi$$^{2}$-test resoundingly rejects the hypothesis that metallicity 
is constant with radius ($\chi^{2}$=225 for 17 degrees of freedom). A weighted linear fit to these data confirm a metallicity gradient of -0.075$\pm$0.011
dex deg.$^{-1}$ (black dashed line). As is evident from an inspection of figure~\ref{metrad}, this is less than the radial variation inferred from the 
[Fe/H] measurements of \cite{carrera08} for field RGB stars, which drop to [Fe/H]$<$-1.5 beyond 4--5$^{\circ}$. However, the steeper gradient in that study 
might be related to the decrease in the mean magnitude of their RGB star subsamples as a function of increasing galactocentric distance. Having re-appraised 
the measurements of \cite{parisi10}, by calculating the median metallicity of each of their sub-samples (which is more appropriate given the departure of 
the distribution from a Gaussian), their metallicities appear to be in accord with our work. A weighted linear regression to their [Fe/H] data 
(figure~\ref{metrad}, open squares), for the range of radii covered by our 2dF/AAOmega observations, can be also be regarded as consistent with the 
existence of a subtle abundance gradient (-0.046$\pm$0.024 dex deg.$^{-1}$). 

We note that the three outermost (ie. beyond 5$^{\circ}$) RGB star metallicity data points of \cite{parisi10} sit somewhat above the trend delineated by the 
measurements at smaller radial distances. Since the luminosity corrections applied in their metallicity estimations were reliant on them determining $V_{HB}$ from 
rather sparsely populated optical colour-magnitude diagrams, it is conceivable that the abundances for these three fields are systematically in error. As a
check, we have used the near-IR approach followed here to re-appraise the metallicities of the stars in HW\,86, the only one of these fields for which sufficiently 
deep photometry (from IRSF) is available. Assuming $K_{S_{RC}}$=17.35$\pm$0.02, we obtain a slightly smaller estimate of the median metallicity but the difference from
the optically derived value is not significant and does not shift [Fe/H] here onto the linear trend. Thus considering our data and that of \cite{parisi10} together,
there is a suggestion that while the metallicity drops relatively rapidly with increasing radius between 0.5--2-3$^{\circ}$, it appears to be almost constant 
beyond 3--4$^{\circ}$.  Interestingly, this is comparable to the radial distance at which the RGB star radial velocities change sign along position angle,
 PA$\approx$120$^{\circ}$ (Dobbie et al. 2014).

\begin{figure}
\label{agehist}
\includegraphics[angle=0, width=\linewidth]{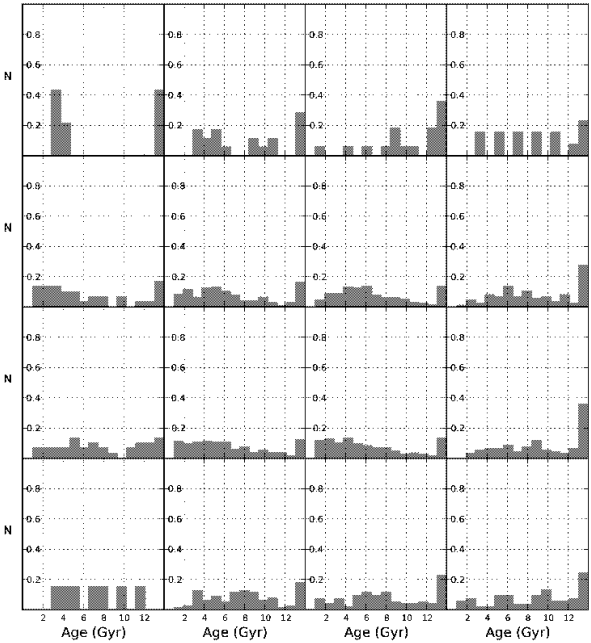}
\caption{Age histograms for the 1865 RGB stars (including a small number observed with more than one field pointing) with $V$ and $I$ photometry as distributed across each of our 
inner 16 2dF field pointings.}
\label{agehist}
\end{figure}

\section{Interpretation and discussion}

\subsection{Metallicity and age across the SMC}

Intrigued by the trends identified above and in view of the large number of stars and the extensive area over which we have abundance measurements, we have  
constructed a metallicity surface for our survey area by estimating this parameter at a series of regularly spaced grid points in RA and declination (every 
10 arcmin), using a bi-variate gaussian smoothing kernel, with an adaptive width corresponding to one third the distance to the 200th closest star, to weight the
individual measurements \citep[e.g.][]{walker06}. A contour plot of this surface is displayed in figure~\ref{metmap} and a map of the smoothing kernel width is shown in figure~\ref{metmapkern}. To first order, these contours are circular 
and centered on the Cloud, re-affirming the prevalance of lower metallicities at larger radial distances. The [Fe/H]=-1.0 contour neatly encapsulates the optically 
bright regions of the SMC and the metallicity appears to change most rapidly at around these radial distances. The plot suggests that the drop-off in metallicity with 
increasing distance from the center is less pronounced towards the limits of our survey region, but the composition gradient also seems to be smaller within the inner degree.  We do not believe these features are due merely to the smoothing kernel applied here since a similar plot constructed for a synthetic version of these data, where the decrease in metallicity with radius has been assumed to be perfectly linear, fails to reproduce them.
This latter feature could be due to our survey being insensitive to the stellar populations formed within these central regions within the last 1Gyr and 
which are likely to be the most metal rich SMC stars \citep[e.g.][]{hill97}.

Subsequently, to explore the star formation history across the Cloud, we have also estimated the ages of our red-giant stars, taking advantage of our spectroscopic 
abundance measurements to break the degeneracy, with respect to age and [Fe/H], of their locations in colour-magnitude space. It is worth re-iterating here the point 
raised in \cite{cole05} that calculations of this nature are prone to substantial uncertainties and as a consequence do not provide a precise evaluation of the age of any
given star. For example, a small number of our red-giants are likely to be asymptotic giant branch stars \citep{cole00}. In these cases the assumption that the 
stars are on the RGB results in roughly a 30\% underestimation of their ages. Moreover, these estimates can be impaired by subtle variations in the reddening across 
the SMC and the three dimensional structure of the Cloud. \cite{zaritsky02} found the extinction distribution for the intermediate age population has a long tail
extending to A$_{V}$$\approx$1, while the finite depth of the SMC amounts to $\Delta$ $m$$\approx$$\pm$0.2 mag. \citep{subramanian12}. Nevertheless, the generous size 
of our sample can help to mitigate the impact of these uncertainties.

Although moderately deep, uniform, near-IR photometry is available for our entire survey area from the 2MASS-6X survey \citep{skrutskie06}, we have sacrificed some
of our extensive spatial coverage for accurate $V$ and $I$ band photometry from the Magellanic Clouds Photometric Survey \citep[MCPS][]{zaritsky02}. The $V$-$I$ colour
provides a better handle (by a factor of approximately three) on the effective temperatures, and thus masses and ages, of the stars than the $J$-$K$ colour
\citep[e.g.][]{bessell98}. Additionally, with the optical photometry we can obtain an estimate of the age and its corresponding uncertainty by exploiting the relation 
\cite{carrera08a} have determined between $log_{10}$ (age) and $M_{V}$-$M_{I}$, M$_{V}$ and [Fe/H]. From cross-correlating our red-giant sample against the MCPS dataset
we have obtained $V$ and $I$ band photometry for 2362 observed sources. Subsequently, we have determined M$_{V}$ and $M_{V}$-$M_{I}$ for 2185 of these, accounting for the 
effects of reddening using estimates of this from the extinction catalogue of \cite{haschke11}. This work is based on photometry of red clump and RR Lyrae stars from 
the third data release from the Optical Gravitational Lensing Experiment \citep[OGLE, ][]{udalski03} and unfortunately does not cover the entire area of the MCPS. 
Additionally, we have assumed a distance modulus for the SMC of ($m$-$M$)$_{0}$=18.90 \citep[e.g.][]{cole98,udalski98,twarog99,kapakos11,haschke12}. 

Our age estimates for the stars within the 16 innermost 2dF field pointing are shown plotted against [Fe/H] in 
figure~\ref{agemetpt}. The uncertainties in age listed in Table~\ref{ages} have been estimated following a Monte-Carlo approach.  We generated thousands of synthetic 
versions of each star, randomly drawing the magnitudes and the metallicity from normal distributions centered on the measured values of $V$, $I$ and [Fe/H] and with widths 
corresponding to the estimated errors on these parameters. Subsequently, these synthetic values of $V$, $I$ and [Fe/H] were propagated through the age-colour-magitude-metallicity
relation of \cite{carrera08a}. We adopted values for the co-efficients that were similarly drawn from normal distributions centered on their most likely values and with widths
corresponding to the quoted  co-efficient  errors. Where the uncertainties have scattered a calculated age to an unphysically large value (ie. greater than that of the Universe), we have followed
a similar approach to previous investigators \citep[e.g.][]{carrera08} and reset this to 13.5Gyr \citep[ie. based on a 13.7Gyr old Universe, ][] {spergel03}. Since RGB stars younger than 1Gyr are not expected to lie near our initial colour-magnitude selection box, the proportion of objects which are determined to be less than 1Gyr old likely 
suffer from signficant measurement error or are potentially AGB or RSG star interlopers. Consequently, for our subsequent analysis we have clipped the sample at an age of 1Gyr, leaving us age/metallicity 
estimates for 1835 unique sources. We do not expect this approach to dramatically affect our conclusions since our comparison of ages and metallicities from field to field is not of a detailed nature. If required by other investigators, age estimates for all 2185 objects with $V$ and $I$ photometry and reddening data can be re-constructed from the information in Tables~\ref{rgstars} and ~\ref{ages}.

\begin{figure}
\label{agemetne}
\includegraphics[angle=0, width=\linewidth]{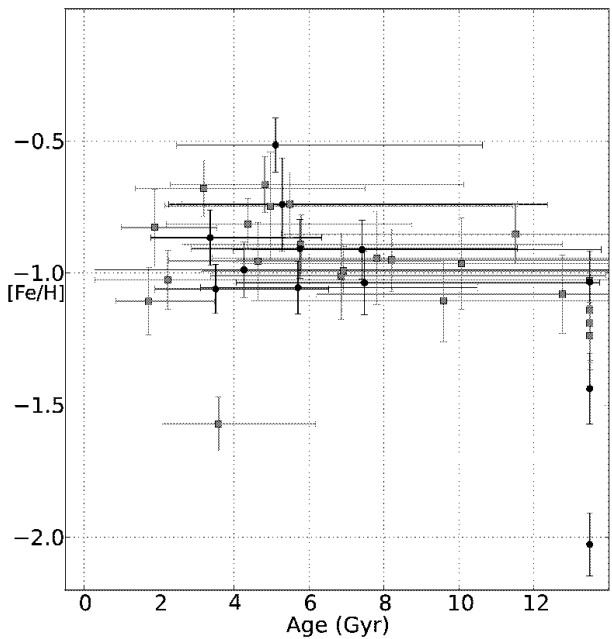}
\caption{Age-metallicity data for the two sub-samples of stars in the NE region of our survey, $v_{r}$$\le$140kms$^{-1}$ (black dots) and $v_{r}$$>$140kms$^{-1}$ (gray squares).}
\label{agemetne}
\end{figure}

% use plotting code in Publications/SMC directory
% original datafiles in GIRARDI/AUTOG 

The data shown in figure~\ref{agemetpt} affirm the previously identified trend between metallicity and age in the SMC where-by the most metal-rich stars tend to be the youngest \citep[][]{dacosta98, idiart07,carrera08,cignoni13}. For example, in his analysis of the ages and metallicities of dozens of SMC star clusters, \cite{piatti11} found a decrease from around [Fe/H]$\approx$-0.5 -- -1.0 for the 1-2 Gyr old populations down to typically [Fe/H]$\approx$-1.0 -- -1.5 for the populations older than 5-6Gyr. Our data are also accordant with an age-metallicity relation that does not change form substantially across 
the region covered by our survey. In light of this, we concur with \cite{carrera08} that the observed radial variations in the metallicity are attributable to differences 
in the proportions of stars of different ages across the SMC, with younger populations concentrated in the central regions. For example, stellar age distributions, 
constructed for the stars in each of the 16 innermost 2dF field pointings (figure~\ref{agehist}), show a steady increase to $\tau\simless$4Gyr in the central regions of the
Cloud, whereas in the surrounding fields these tend to be flatter. This finding is also accordant with the results from recent deep optical imaging studies of the star 
formation history in several small fields distributed across the SMC \citep[e.g.][]{dolphin01, cignoni12}. Based on their population synthesis analyses of 
colour-magnitude diagrams \cite{cignoni13} find that unlike the two fields in their sample located at comparatively large radial distances (greater than two degrees) from 
the center 
of the Cloud, where initially rather modest rates of star formation appear to have ramped down further within the last 5Gyr, the more central fields have experienced relatively 
strong bursts of star formation during this latter period of time.

Notably there are hints of a larger proportion of younger stars in the two fields 3D8 and 3D10, located in the NE of our survey region and in which we see evidence of 
a bimodal velocity distribution (Dobbie et al. 2014). This bimodality appears to be associated with two enhancements in the space densities of intermediate age stars located at
 around 55 and 65 
kpc along these lines of sight \citep{nidever13,hatzidimitriou93}. It has been proposed by \citeauthor{nidever13} that the former is a stellar analogue of the Magellanic Bridge. 
In view of this we have split the RGB star populations of each of these fields into two on the basis of their radial velocities (at 140kms$^{-1}$). After applying a small correction to 
the metallicity measurements of the lower velocity stars to account for their smaller distances ($\Delta$[Fe/H]=+0.06), we have examined the metallicity and the age distribution 
for each velocity sub-sample. K-S tests of these age and metallicity datasets ($P=0.70$ for metallicity and $P=0.64$ for age) are unable to distinguish the two sub-samples.
Moreover, the age-metallicity relations for these sub-samples (figure~\ref{agemetne}) appear to be very similar in form and thus support the proposal of \citeauthor{nidever13} 
that the stellar population found around 10kpc closer to the Sun originated from the SMC, having possibly been stripped off during the most recent encounter with the LMC. Indeed,
our crude estimate that these structures separated around 300Myr, which we derived from the relative position of the peaks in the radial velocity histograms for fields 3D8 and 
3D10 (approximately 120 kms$^{-1}$ and 155kms$^{-1}$, see Dobbie et al. 2014) and the line-of-sight distance between them, is comparable to the time of the last close encounter 
predicted by back integration of the Clouds' orbits \citep[200Myr, e.g.][]{bekki09}.

While the substantial proportion of younger stars in the foreground structure could merely be down to low number statistics, it might also indicate 
that the most recent encounter between the two Clouds has gouged out stars from the inner regions of the SMC and not merely stripped older metal poor
stars from the galaxy's periphery.  Indeed, the removal of stars from the depths of the SMC's potential well is envisaged by a recent simulation where
the last encounter between the Magellanic Clouds entailed their direct collision \citep{besla12}. This model, which naturally accounts for some of the 
more distinctive morphological characteristics of the LMC (e.g. the offset of the photometric and kinematic centers), also predicts that a significant 
population of stars torn from the SMC is accreted onto the LMC. \cite{olsen11} recently unearthed 30 AGB stars towards the LMC which, from consideration 
of their kinematics and metallicities, are believed to have originated in the SMC. While the differences we observe between the population densities
of the Bridge and the Counter-Bridge structures identified kinematically in Dobbie et al. (2014), also seem to concur with the predictions of the \cite{besla12} 
maximum interaction model, the median metallicity of the accreted SMC stars is only [Fe/H]= -1.25$\pm$0.13. This is comparable to the values we observe
in the outer parts of our study (e.g. [Fe/H]=-1.1-- -1.2, see Table~\ref{Metals}), arguing that these objects are predominantly from the periphery
of the Cloud. This perhaps indicates that the most appropriate description of the interaction history of the Magellanic system lies somewhere between that 
defined by a close approach simulation \citep[e.g. ][ model 1]{diaz12, besla12} and a direct collision model \citep[e.g. ][ model 2]{besla12}. 

\label{s6}

\subsection{Star formation in the SMC}

Our RGB star abundance and age determinations and those from other recent studies of the SMC are indicative of an outside-in progression of star formation activity in this 
galaxy \cite[e.g.][]{carrera08,noel09}. A similar pattern has been identified in a number of other low luminosity galaxies, including dwarf irregulars, which have been subjected to 
analyses of their star formation histories as a function of galacto-centric radius. For example, \cite{gallart08} studied deep colour-magnitude diagrams of four fields at a 
range of galacto-centric radii (r$\approx$2--7$^{\circ}$) in the LMC and found the age of the youngest stellar population to increase progressively from presently active star 
formation to 1.5Gyr, stepping further out from the center. Additionally, a deep imaging investigation of four isolated dwarf spheroidal and transition galaxies finds that while 
a 13 Gyr old generation of stars is present throughout each, in three out of the four systems (Tucana, LGS-3 and Phoenix), the radial exponential scalelengths of the stellar 
distributions decrease with time \citep{hidalgo13}. 

These low luminosity dwarfs do not appear to behave merely as scaled down versions of large galaxies, like the Milky Way and M33, where, in accordance with the predictions 
of $\Lambda$-CDM cosmology, it has been found that star formation activity is propagating outwards in their disks \citep[][]{yuan11, williams09g}. Nevertheless, in recent 
hydrodynamical simulations of the evolution of isolated dwarf galaxies with masses comparable to the Magellanic Clouds, stars form initially in both the central regions and in
widely distributed local condensations, rapidly enriching the primordial gas with metals. As the gas is consumed faster in the denser inner parts of these galaxies where the 
initial star formation rates are higher, the reduction in pressure here draws newly enriched gas inwards and the star formation retreats towards their centers \citep{pilkington12}.
It is also anticipated that as the result of shocks between colliding supernovae driven outflows and inflowing gas, some stars still form at larger galacto-centric distances 
\citep{stinson09},  but as a consequence of the overall inwards trend, the lower metallicity stars will have a more extended distribution at the present epoch, as is observed in 
the SMC. 

The cycle of periods of increased star formation activity followed by SNe outflows and stellar winds reducing the fuel available for further genesis of new stars by driving enriched
gas out of the galaxies' central regions \cite[e.g.][]{governato10}, is envisaged to lead to 50\% of the stellar mass being in place after only 4Gyr of evolution in a typical dwarf 
system \citep[][]{shen13}. This rate of growth of stellar mass is consistent with empirical estimates for many of the irregulars in the ANGST dwarf galaxy sample of \cite{weisz11}.
However, the SMC is somewhat atypical in that it appears to have experienced relatively modest rates of star formation within the first 6-7 Gyr, as evidenced, for example, by the 
relative rarity 
of horizontal branch stars \citep[][]{noel07}. Deep colour-magnitude diagrams for the Cloud confirm a fitful history of star formation but where the substantial increases in activity have 
occured within the last 5-6Gyr \citep[e.g. 4-6Gyr and 2-3Gyr ago][]{noel09,cignoni12}. Enhancments around 5-6Gyr and 2Gyr ago are also evident in the formation rate of SMC star clusters \cite{piatti11}. The age distribution of the stars drawn from the most metal rich quartile
of our SMC red-giants is largely restricted to values less than 6Gyr, whereas the age distribution of the corresponding metal poor stars indicates the bulk of these have formed 
prior to this period (figure~\ref{agecum}). Interestingly, this suggests an association between the bursts of star formation activity within the last 5-6Gyr which are confined to or at least 
significantly more intense in the inner regions of the SMC and the kinematical signature we observe in our metal rich RGB stars, which we tentatively ascribed to a disk-like structure (Dobbie et al. 2014). 

%No metallicity gradient observed in old RR lyrae population \cite{haschke12}

Although recent hydrodynamical simulations can reproduce outside-in progression of star formation in isolated low mass galaxies, the slightly unusual star formation history of the 
SMC suggests another mechanism may have driven or at least be influencing the star formation rate here. Two mechanisms which can enhance this in galaxies and are 
relevant in the context of the SMC are mergers \citep[e.g.][]{mihos94,sanders88} and tidal interactions \citep[e.g.][]{bekki08}. \cite{hopkins09} highlight that from both the 
theoretical and the empirical points of view, almost every galaxy has undergone a minor merger (mass ratio $\simless$0.1) within the last few billion years and that major mergers are
also not uncommon. In the merging of two gas rich galaxies, torques between stellar and gaseous structures arising from the non-axisymmetric potentials result in a proportion of the 
gas losing its angular momentum and falling into central regions. This leads to a substantial upturn in the star formation rate here and the pre-merger stars being transformed into a
pressure supported population \citep{hopkins13}. A number of investigators have attributed the quite different spatial and kinematic properties of the gas, the young stars and the 
red-giants in the SMC to a merger several billion years ago \citep[e.g.][]{rafelski05,bekki08,subramanian12}. However, \cite{stinson09} have demonstrated that processes such as disk 
sloshing and simple disk shrinkage can lead to the formation of extensive halo-like stellar populations around dwarf galaxies, without the need to invoke galaxy-galaxy mergers 
\citep[see also ][]{nidever11}. A recent theoretical exploration of the chemical evolution of the SMC, which treats a galaxy-galaxy merger several Gyr ago with a mass ratio 
of between 1:1 to 1:4, anticipates a rather pronounced dip in the age/metallicity relation around this time \citep{tsujimoto09}. There seems to be little support for an event of this 
magnitude from recent empirical determinations based on both star clusters and field stars \citep{piatti11,piatti12}. Moreover, the lack of a large difference between the velocity 
dispersions of the metal rich ($\sigma_{v_{los}}$$\approx$22kms$^{-1}$) and the metal poor ($\sigma_{v_{los}}$$\approx$26kms$^{-1}$) quartiles of our RGB star sample, predominantly formed 
less than and more than 6Gyr ago, respectively, also seems to disfavour the notion that a major merger occurred prior to 4-5Gyr ago. It remains possible that systems with more 
extreme mass ratios have merged with the SMC. Indeed, the subtle positive correlation of the velocity dispersion of our RGB star population with age (Dobbie et al. 2014) is in accord
with the overall trend predicted by recent cosmological dwarf galaxy simulations of \cite{shen13}, which they attribute to the scattering of the stars that formed early-on by 
(presumably minor) mergers. 

\begin{figure}
\label{agecum}
\includegraphics[angle=0, width=\linewidth]{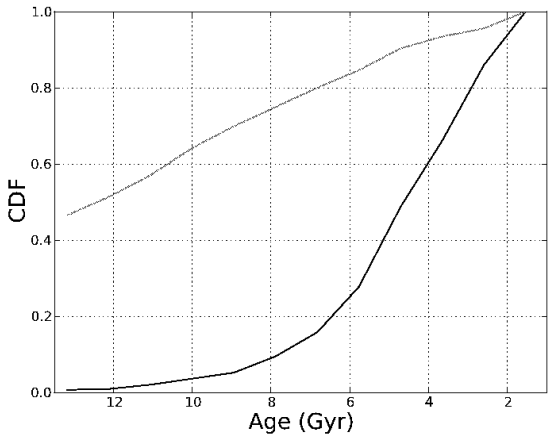}
\caption{Cumulative distributions for the ages for those sources within the metal rich (black line) and metal poor (gray line) quartiles of our RGB star sample for which we have MCPS $V$ and $I$ band photometry and
age estimates.}
\label{agecum}
\end{figure}

\cite{cignoni12} have proposed that tidal interactions between the Magellanic Clouds and the Galaxy about 5.5 Gyr ago might be responsible for the first substantial
burst, which has also been associated with formation of the LMC Bar \citep[e.g.][]{smecker02}. Tidal interactions can precipitate the formation of dynamical structures which drive 
gas inwards and stimulate star formation, particularly within the central regions of galaxies through the compression of gas to critical densities necessary to ignite the process 
\citep{hopkins09}.
In a recent theoretical investigation of the relationships amongst these three galaxies, where the proper motion of the SMC has been derived from a combination of ground-based 
\citep{vieira10} and space-based measurements \citep[][]{kallivayalil06}, \cite{diaz12} find that for at least the last 5Gyr, the Magellanic Clouds have orbited within the virial 
radius of the Galaxy and have experienced three peri-Galactic passages of around 50kpc, including one about 5Gyr ago. Since the LMC and SMC were separated at this time by more than 
100kpc, it is unlikely, in this interpretation, that tidal interaction between the Clouds triggered the burst of activity imprinted on the star formation histories of both galaxies 
4-6Gyr ago. However, cosmological studies of satellite populations of Milky-Way like galaxies and a very recent assessment of the Clouds proper motions have cast doubt on whether the
SMC-LMC system has previously experienced a close encounter with the Galaxy \citep[][]{tollerud12,kallivayalil13}. The preferred ``first orbit'' scenario, makes it difficult to 
attribute the increases in the SMC star formation activity at 4-6Gyr and 2-3Gyr ago to peri-Galactic passages. The Magellanic Stream is then explained as forming around 2Gyr ago when
the LMC and SMC first became a strongly coupled binary system \citep[e.g.][]{diaz12}. 

The coincidence between the times of major star formation episodes around 5Gyr (and at 2-3Gyr) in both Magellanic galaxies points towards a common trigger. If this was not a tidal 
encounter with the Milky Way then it seems likely to have been a mutual interaction. Indeed, in the orbital models of \cite{besla10}, the LMC and SMC experienced a pericenter passage around 5Gyr 
ago when they were separated by only 30-40 kpc. The subsequent LMC-SMC pericenter also appears to coincide with another Magellanic wide burst of star formation around 2-3Gyr ago 
\citep[e.g.][]{harris09}, when the Stream is believed to have formed, although these models are unable to reproduce the current positions and velocities of the Clouds. Nonetheless, 
considering the cosmological arguments for them being on their first orbit of the Galaxy and the results of recent theoretical and observational investigations, including the star 
formation histories of both galaxies, the results of our kinematical investigation of the SMC RGB star population (Dobbie et al. 2014) and the stellar abundance and age estimates, we
are led to favour a model in which interactions between the SMC and the LMC, beginning 5-6 Gyr ago, led to an increase in the flow of gas into the central regions of the nascent disk 
of the former that promoted the development of giant molecular clouds and resulted in an upsurge in the star formation rate. However, until better constraints are available on the 
proper motions of both Magellanic Clouds it is unlikely that it will be possible to reach a definitive conclusion on this matter.

\section{Summary}

We have obtained optical spectroscopy for 3037 SMC red giant stars distributed across 37.5 deg$^{2}$ of the SMC. From 
these data we have measured their metallicities and have: 

\begin{itemize}

\item{affirmed that the median metallicity of stars at larger radial distances from the optical center of the Cloud is lower, at least out to $r$$\approx$5$^{\circ}$. A linear
model fit to [Fe/H] as a function of radial distance has a gradient of -0.075$\pm$0.011 dex deg$^{-1}$.} 

\item{exploited the MCPS data to estimate ages for 1835 RGB stars, confirming that the stellar age-metallicity relation has a similar form 
across our survey area and that there is a trend in the SMC for younger objects to have larger metallicities. The observed radial 
abundance gradient reflects that stars in the central regions of the SMC are generally younger than those at larger radial distances.}

\item{demonstrated that the age-metallicity relation of the intermediate age stellar population located around 10kpc in front of the NE of the 
Cloud is similar to that of the main body of the SMC, in accord with the recent proposal by \citeauthor{nidever13} that this is a 
stellar analogue of the Magellanic Bridge.}

\item{found that the stars in the spatially extended most metal poor quartile of our RGB sample are typically older than 6Gyr.}
 
\item{determined that the metal rich quartile of our RGB stars predominantly have ages less than approximately 6Gyr, and were probably formed during periods of substantial enhancements in the star formation rate of the SMC. These appear to be associated with bulk rotational motion which we have tentatively associated with disk-like structure.}

\item{after consideration of recent theoretical and observational investigations of the Clouds and cosmological arguments, concluded that 
an interaction between the LMC and the SMC was the likely trigger for the upturn in stellar genesis that is imprinted on the star formation 
history of the central SMC around 5-6Gyr ago.}  

\end{itemize}

\section*{Acknowledgments}

This publication makes use of data products from the Two Micron All Sky Survey, which is a joint project of the 
University of Massachusetts and the Infrared Processing and Analysis Center/California Institute of Technology, 
funded by the National Aeronautics and Space Administration and the National Science Foundation. This research was
supported under the Australian Goverment's Australia-India Strategic Research funding scheme (reference number ST040124).
We thank Russell Cannon for his input during the planning phases of this project. We thank the referee for helpful and 
constructive reports on our two SMC papers. 

%%%%%%%%%%%%%%%%%%%%%%%%%%%%%%%%%%%%%%%%%%
%%%%%%%%  Bibliography  %%%%%%%%
%%%%%%%%%%%%%%%%%%%%%%%%%%%%%%%%%%%%%%%%%%
%
\bibliographystyle{mn2e}
\bibliography{mnemonic,therefs}

\bsp

\label{lastpage}

\end{document}